\documentclass[aps,prx,reprint,superscriptaddress]{revtex4-1}
\usepackage{graphicx, color}
\usepackage{amsfonts,amssymb,amsmath}
\usepackage{siunitx}

\begin{document}

\title{Novel self-epitaxy for inducing superconductivity in the topological insulator (Bi$_{1-x}$Sb$_x$)$_2$Te$_3$}

\author{Mengmeng Bai}
\thanks{These authors contributed equally}
\affiliation{Physics Institute II, University of Cologne, Z\"ulpicher Str. 77, 50937 K\"oln, Germany}

\author{Fan Yang}
\thanks{These authors contributed equally}
\email[]{fanyangphys@tju.edu.cn}
\affiliation{Physics Institute II, University of Cologne, Z\"ulpicher Str. 77, 50937 K\"oln, Germany}
\affiliation{Center for Joint Quantum Studies \& Department of Physics, School of Science, Tianjin University, Tianjin 300350, China}

\author{Martina Luysberg}
\thanks{These authors contributed equally}
\affiliation{Ernst Ruska-Centre, Forschungszentrum J\"ulich, 52425 J\"ulich, Germany}

\author{Junya Feng}
\affiliation{Physics Institute II, University of Cologne, Z\"ulpicher Str. 77, 50937 K\"oln, Germany}

\author{Andrea Bliesener}
\affiliation{Physics Institute II, University of Cologne, Z\"ulpicher Str. 77, 50937 K\"oln, Germany}

\author{Gertjan Lippertz}
\affiliation{Physics Institute II, University of Cologne, Z\"ulpicher Str. 77, 50937 K\"oln, Germany}
\affiliation{KU Leuven, Quantum Solid State Physics, Celestijnenlaan 200 D, 3001 Leuven, Belgium}

\author{A. A. Taskin}
\affiliation{Physics Institute II, University of Cologne, Z\"ulpicher Str. 77, 50937 K\"oln, Germany} 

\author{Joachim Mayer}
\affiliation{Ernst Ruska-Centre, Forschungszentrum J\"ulich, 52425 J\"ulich, Germany}

\author{Yoichi Ando}
\email[]{ando@ph2.uni-koeln.de}
\affiliation{Physics Institute II, University of Cologne, Z\"ulpicher Str. 77, 50937 K\"oln, Germany}

\begin{abstract}
Using the superconducting proximity effect for engineering a topological superconducting state in a topological insulator (TI) is a promising route to realize Majorana fermions. However, epitaxial growth of a superconductor on the TI surface to achieve a good proximity effect has been a challenge. We discovered that simply depositing Pd on thin films of the TI material (Bi$_{1-x}$Sb$_x$)$_2$Te$_3$ leads to an epitaxial self-formation of PdTe$_2$ superconductor having the superconducting transition temperature of $\sim$1 K. This self-formed superconductor proximitizes the TI, which is confirmed by the appearance of a supercurrent in Josephson-junction devices made on (Bi$_{1-x}$Sb$_x$)$_2$Te$_3$. This self-epitaxy phenomenon can be conveniently used for fabricating TI-based superconducting nanodevices to address the superconducting proximity effect in TIs. 
\end{abstract}

\maketitle

\section{Introduction}

The seminal proposal by Fu and Kane \cite{Fu2008} to use the superconducting proximity effect for engineering a topological superconducting (TSC) state in a topological insulator (TI) has been extended to various materials platforms \cite{Alicea2012} and pursued by many researchers, with the hope to realize topological quantum computing \cite{Lutchyn2018}. 
The key idea of the Fu-Kane proposal is to induce Cooper pairing among electrons that are effectively ``spinless", which is naturally realized in the TI surface states due to the spin-momentum locking \cite{Sato2017}. Hence, in TIs, the spinless nature can be utilized as long as the chemical potential $\mu$ is located within the bulk band gap which is typically of the order of $\sim$100 meV \cite{Ando2013}. Similar spinless situation can be created in strongly spin-orbit-coupled semiconductors by applying a high magnetic field to remove the states near $\mathbf{k}$ = 0 via the Zeeman-gap opening \cite{Lutchyn2010, Oreg2010}. An important requirement in engineering a TSC state is that those spinless states are proximitized by a superconductor (SC) to develop pair correlations. In this context, epitaxially-grown Al on InAs realizing a highly-transparent interface was a major breakthrough \cite{Krogstrup2015} in the pursuit of engineered TSC state. However, the spinless situation in this case is reached only when $\mu$ is tuned into the Zeeman gap of the order of a few meV, and achieving this condition throughout the sample is difficult in the presence of disorder \cite{Lutchyn2018}. In this respect, TIs could be better suited to the engineering of the TSC state. 

While TIs are in principle a promising platform, preparing a highly-transparent TI/SC interface to induce robust Cooper pairing in TIs has not been straightforward \cite{Veldhorst2012, Oostinga2013, Galletti2014, Snelder2014, Ghatak2018, Jauregui2018, Schuffelgen2019}. In particular, epitaxial growth of a SC layer on a TI to achieve a good proximity effect has never been reported. Since the epitaxial growth of Al on InAs was shown to be crucial for the semiconductor platform in its advancement of the proximity-effect devices \cite{Krogstrup2015, Lutchyn2018}, it is desirable to be able to make similar devices for TIs. 

In our efforts to address this challenge, we discovered that simply depositing Pd on thin films of the TI material (Bi$_{1-x}$Sb$_x$)$_2$Te$_3$ (BST) leads to an epitaxial self-formation of PdTe$_2$ superconductor, which proximitizes the TI and can be used for fabricating Josephson junctions. This will allow for easy fabrication of nano-devices to address proximity-induced superconductivity in TIs and may eventually help to elucidate the Majorana fermions \cite{Alicea2012} that are expected to emerge there.

\begin{figure*}[t]
\centering
\includegraphics[width=11cm]{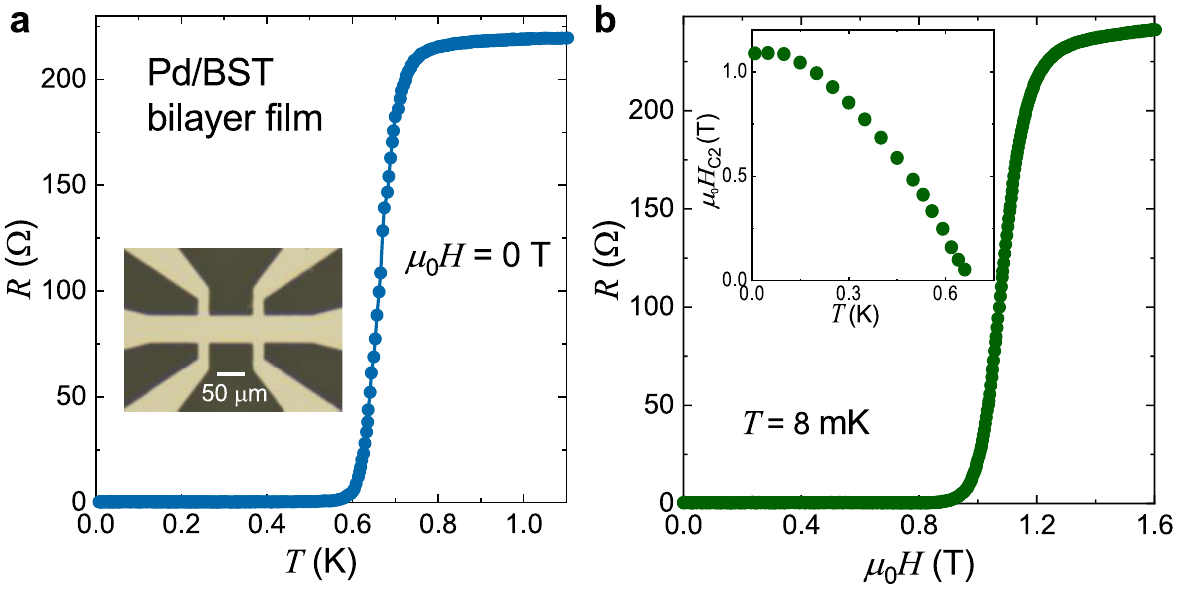}
\caption{Superconducting transition of Pd/(Bi$_{1-x}$Sb$_x$)$_2$Te$_3$ bilayer films (sample 1). (a) Temperature dependence of the resistance ($R$) of a four-terminal device made from sample 1. 
Inset: optical image of the Hall-bar for transport measurement. 
(b) Magnetic-field dependence of $R$, measured at $T$ = 8 mK in perpendicular magnetic field. Inset shows the temperature dependence of the upper critical field $H_{c2}$, defined as the midpoint of the $R(H)$ transition.}
\label{fig:1}
\end{figure*}

\section{Experimental Details}

{\bf BST thin film.}
(Bi$_{1-x}$Sb$_x$)$_2$Te$_3$ films were grown on sapphire (0001) substrates by co-evaporation of high-purity Bi, Sb, and Te from Knudsen cells in the ultra-high vacuum molecular beam epitaxy (MBE) chamber. During the growth, the flux ratio of Bi and Sb was kept at 1:5.5, with which the $x$ value of the grown films was 0.85. 
A two-step deposition procedure was employed, in which the substrate temperature was changed as follows: ramping from 220$^{\circ}$C to 280$^{\circ}$C in 14 min and then keeping it at 280$^{\circ}$C for the time period which is sufficient to grow a film with a desirable thickness. 
The thickness and morphology of the grown films were measured {\it ex situ} by an atomic force microscope (AFM). 

{\bf Pd deposition.}
We employed both magnetron sputtering and thermal deposition for preparing Pd layers, but the bottom portion of the Pd layer to come in contact with BST was always deposited by sputtering. The freshly grown BST films were immediately transferred into a vacuum chamber for Pd sputtering (see \cite{Supplemental} for the characterization of sputtered Pd films). For BST films that have been exposed to air for some time, an additional process of low-power Ar plasma etching (with the power of 10 W for 1 minute) was applied to etch down the film by $\sim$2 nm for the purpose of surface cleaning.  No difference in transport properties was observed between the freshly-grown films and air-exposed-and-cleaned films. 

{\bf EDX and TEM analyses.}
Energy-dispersive X-ray spectroscopy (EDX) analysis and scanning tunneling electron microscopy (STEM) analysis were performed with FEI Titan G2 80-200, which is equipped with a Cs probe corrector (CEOS DCOR), and an in-column Super-X detector for acquisition of energy-dispersive X-ray (EDX) spectra. The EDX line scans, which are averaged along 30 nm parallel to the interfaces, were obtained at an acceleration voltage $V_{\rm acc}$ of 80 kV. High-resolution high-angular dark-field images were recorded at $V_{\rm acc}$ = 200 kV.

{\bf Device fabrication.}
The Hall-bar devices for characterizing the Pd/(Bi$_{1-x}$Sb$_x$)$_2$Te$_3$ bilayer films were patterned with photolithography and Ar plasma etching. To avoid heating which degrades BST films, the photo-resist was cured at room temperature for 12 hours without baking. Josephson junctions and a superconducting quantum interference device (SQUID) were fabricated using electron-beam lithography. After the resist development, samples were cleaned in Ar plasma at 10 W for 1 minute to remove possible resist residues in the contact areas. The Pd deposition was performed in two steps. First a thin layer of Pd was deposited by magnetron sputtering, and then additional Pd was deposited by thermal evaporation to increase the total thickness of the Pd electrodes; this was necessary for assuring good contrast under SEM for the characterization of the device dimensions. The lift-off was done in warm acetone. 

{\bf Transport Measurement.}
The Hall-bar devices were measured in the four-terminal configuration using a low-frequency ac lock-in technique; all the devices were measured down to 280 mK in a $^3$He cryostat, and the device made from sample 1 \cite{Supplemental} was additionally measured in a dry dilution refrigerator (Oxford Instruments TRITON 200) down to the base temperature of 8 mK. The measurements of the Josephson junctions and the SQUID were performed in the dilution refrigerator in a pseudo-four-terminal geometry. We performed both dc $I$-$V$ and ac differential resistance (d$V$/d$I$) measurements. The superconducting magnet was driven with a Keithley 2450 Sourcemeter to ensure a precise control of the small applied magnetic field with low noise.

\begin{figure*}[t]
\centering
\includegraphics[width=16cm]{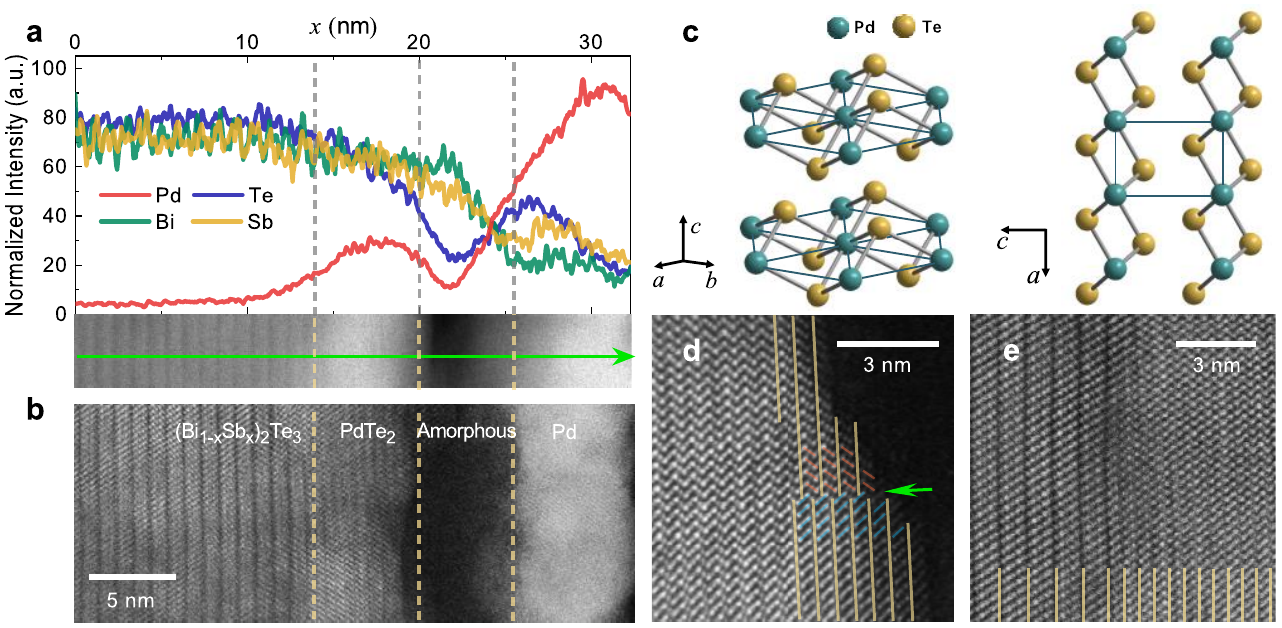}
\caption{EDX and STEM analysis of Pd/(Bi$_{1-x}$Sb$_x$)$_2$Te$_3$ bilayer film (sample 1). (a) EDX line-scan spectra across the interface (upper panel) and a high-angle annular dark-field STEM image of the cross section (lower panel), measured at $V_{\rm acc}$ = 80 kV. The EDX scanning line is shown by the green arrow in the lower panel. (b) Dark-field STEM image taken at $V_{\rm acc}$ = 200 kV. An amorphous layer poor in Pd and Te and a crystalline layer of PdTe$_2$ are observed between the sputtered Pd layer and the (Bi$_{1-x}$Sb$_x$)$_2$Te$_3$ film. (c) Crystal structure of PdTe$_2$; thin turquoise lines in the left (right) panel emphasizes the hexagonal structure (unit cell) formed by Pd. (d) High-resolution STEM image of the PdTe$_2$ layer, taken at $V_{\rm acc}$ = 200 kV. Clear triple-layers with different orientations are observed in the thin parts, as illustrated by the red and blue slash lines. The green arrow indicates a domain boundary. In thick areas zigzag patterns are seen due to the stacking of different domains in the thickness direction. (e) High-resolution STEM image across the PdTe$_2$ /(Bi$_{1-x}$Sb$_x$)$_2$Te$_3$ interface, showing clear evidence of epitaxial growth. All data are obtained from the same Pd/(Bi$_{1-x}$Sb$_x$)$_2$Te$_3$ bilayer film (sample 1) from which the transport results in Fig. 1 were obtained.   
}
\label{fig:2}
\end{figure*}

\section{Results and Discussions}

\subsection{Self-formation of PdTe$_2$ superconductor}

Our discovery of the epitaxial self-formation of PdTe$_2$ superconductor in the Pd-deposited (Bi$_{1-x}$Sb$_x$)$_2$Te$_3$ (BST) surface was serendipitously made while we were trying to find a suitable electrode material for BST that does not require a wetting layer. Since we occasionally observed traces of superconductivity in BST devices having Pd electrodes, we tried to nail it down by preparing Pd/BST bilayer films by sputtering Pd on MBE-grown BST thin films with $x$ = 0.85, which usually gives us bulk-insulating samples \cite{Taskin2017}. The bilayer films were then patterned into a Hall-bar shape using photolithography. As shown in Fig. 1a, relatively sharp superconducting transition was observed in such a device with the midpoint $T_c$ of 0.67 K (transition width $\Delta T_c \simeq$ 0.13 K). The midpoint upper critical field $H_{c2}$ measured on this sample at 8 mK in perpendicular field was 1.09 T (Fig. 1b). Clearly, a superconducting phase is present in this bilayer film, even though neither Pd nor BST is a superconductor. Hence, the observed superconductivity must originate from a new phase self-formed at the interface at room temperature. 

To identify the SC phase in our samples, we employed scanning transmission electron microscopy (STEM) and energy-dispersive X-ray spectroscopy (EDX). Two new phases were observed at the interface: an amorphous layer poor in Pd and Te and a crystalline layer rich in Pd, as shown in Figs. 2a and 2b. The high-resolution STEM image (Fig. 2d) shows that the crystalline phase comprises a triple-layer structure. As the atoms within a triple-layer show different scattering intensities, at least two types of atoms are present.
After a careful comparison of the crystal structure and the lattice constants with known materials, this crystalline phase was identified to be PdTe$_2$, which is composed of hexagonal Te-Pd-Te triple-layers stacked by the van der Waals force along the $c$ direction (Fig. 2c, left) with lattice constants $a$ = 4.04 \si{\angstrom} and $c$ = 5.13 \si{\angstrom}  \cite{Liu2018}. When viewed along the $b$ direction (Fig. 2c, right), the structure matches what is seen in Fig. 2d. 
The observed lattice constants were $a$ = 4.07 \si{\angstrom} and $c$ = 5.44 \si{\angstrom}, pointing to a lattice expansion of 6.0\% in the $c$ direction. The origin of this expansion is most likely the intercalation of Bi/Sb atoms into the van der Waals gap \cite{Supplemental}.

An important information we obtain from the STEM image is that the PdTe$_2$ layer is epitaxially formed with an atomically-sharp interface to BST; in fact, in Fig. 2e one can see clear signatures of van-der-Waals epitaxy. The mechanism of this epitaxial self-formation of PdTe$_2$ that proceeds at room temperature is an interesting topic of future research. We speculate that it is triggered by the diffusion of Pd atoms into BST, resulting in a phase separation into (i) crystalline PdTe$_2$ phase with some Bi/Sb atoms incorporated as intercalant and (ii) amorphous phase containing the Bi/Sb atoms that were pushed out from the PdTe$_2$ phase. This scenario is the most consistent with our EDX result shown in Fig. 2a.

PdTe$_2$ is long known \cite{Guggenheim1961} to superconduct with $T_c$ up to 2.0 K. Recently, PdTe$_2$ was found to be a type-II Dirac semimetal \cite{Noh2017,Fei2017} and was furthermore proposed to realize topological superconductivity in the bulk \cite{Noh2017,Fei2017,Leng2017}. However, very recent experiments show that PdTe$_2$ is most likely a conventional $s$-wave superconductor \cite{Das2018,Voerman2019}. Hence, we can conceive the PdTe$_2$ layer to be the proximitizing SC which brings $s$-wave Cooper pairing into BST.
In fact, the $H_{c2}$ value of 1.09 T measured at 8 mK agrees well with the theoretical zero-temperature value of 1.13 T calculated from the slope $($d$H_{c2}$/d$T)_{T = T_c}$ = $-2.44$ T/K using the Werthamer-Helfand-Hoenberg (WHH)  theory \cite{Werthamer1966}, pointing to a conventional nature of the superconductivity.

\begin{table*}
\begin{tabular}{cccccccc}
 \hline
 \hline
Name & BST layer & Pd (sputt.)& Pd (evap.) &Ar-cleaning & $T_\textrm{c}$ & $\Delta T_\textrm{c}$ & $\mu_{0}H_\textrm{c2}$ \\
 \hline
 \hline
 Sample 1& 68 nm& 8 nm& 0& Yes& 0.67 K& 0.13 K& 0.84 T \\
 \hline
 Sample 2& 50 nm& 20 nm& 0& No& 1.00 K& 0.04 K& 0.34 T \\
 \hline
 Sample 3& 16 nm& 8 nm& 0& Yes& 1.22 K& 0.04 K& 0.48 T \\
 \hline
 Device A& 22 nm& 15 nm& 15 nm& Yes& 0.90 K& -& - \\
 \hline
 Device B& 35 nm& 20 nm& 15 nm& Yes& 0.70 K& -& - \\
 \hline
 \hline
\end{tabular}
\caption{Summary of the preparation parameters and properties of the samples and devices measured in this study. $T_\textrm{c}$ of samples 1 -- 3 is defined as the temperature at the midpoint of the resistive transition. For devices A and B, $T_c$ is defined as the midpoint of the main resistance drop in the $R(T)$ curve.  $H_{\rm c2}$ shown here is defined as the midpoint in the resistive transition measured at $T$ = 280 mK. $\Delta T_c$ is the temperature interval in which the resistance changed from 90\% to 10\% of the normal-state resistance during the superconducting transition. ``sputt.'' and ``evap.'' stand for sputtered and evaporated, respectively. 
}
\end{table*}

We have investigated three differently-prepared samples of Pd/BST bilayer films by fabricating Hall-bars. All samples, regardless of the details of the Pd deposition process, showed robust superconductivity with $T_c$ = 0.67 -- 1.22 K, as summarized in Table I. These $T_c$ values are lower than the $T_c$ of bulk PdTe$_2$, which is probably due to the limited thickness of the PdTe$_2$ layer ($< 10$ nm). Such a reduction in $T_c$ was also reported for PdTe$_2$ thin films grown by molecular-beam epitaxy (MBE) \cite{Liu2018}.

\subsection{Josephson junction}

\begin{figure*}[t]
\centering
\includegraphics[width=11cm]{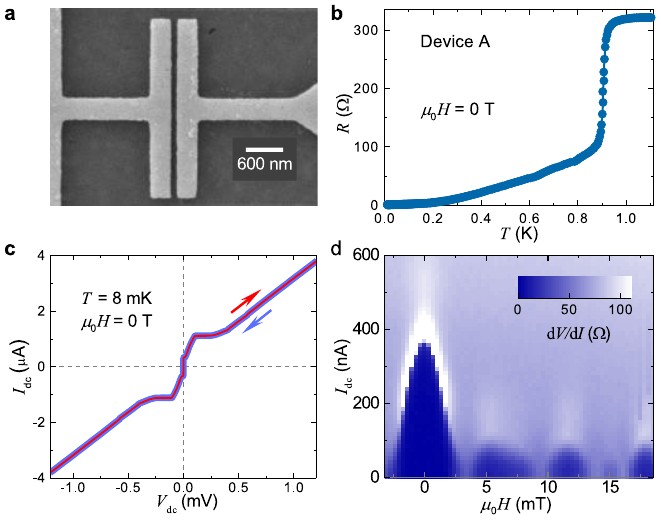}
\caption{Josephson junction fabricated by Pd deposition. (a) SEM image of the Josephson junction device (device A) with two parallel Pd electrodes deposited on top of a 23-nm-thick (Bi$_{1-x}$Sb$_x$)$_2$Te$_3$ film ($x$ = 0.85). The length and width of the junction channel are $\sim$100 nm and 3 $\mu$m, respectively. (b) Zero-bias resistance ($R$) of device A as a function of temperature in $H$ = 0. (c) $I$-$V$  characteristics of device A, measured by scanning $I_{\rm dc}$ in both positive (red curve) and negative (blue curve) directions at $T$ = 8 mK in $H$ = 0. No hysteresis was observed. (d) 2D mapping of the deferential resistance (d$V$/d$I$) in the $I_{\rm dc}$ vs $H$ plane, measured at $T$ = 8 mK in device A.  
}
\label{fig:3}
\end{figure*}

To see if this self-formation of the SC phase can be used for actually making TI-based superconducting devices, we fabricated a Josephson junction and a SQUID by defining the superconducting parts via electron-beam lithography and Pd deposition/lift-off (see Table I for important parameters). A scanning electron microscope (SEM) image of the Josephson junction (device A) is shown in Fig. 3a. The separation between the Pd electrodes in this device was $\sim$100 nm. From the resistance ($R$) vs temperature ($T$) curve shown in Fig. 3b, one can see that the self-formed PdTe$_2$ layer beneath the Pd electrode in this device had $T_c \approx$ 0.90 K and the Josephson coupling between the two SC electrodes is established to give zero resistivity at $T \lesssim$ 150 mK. This device is designed to have narrow ($\sim$500 nm) SC electrodes to minimize flux jumps which disturb Fraunhofer-pattern measurements \cite{Ghatak2018}.

The $I$-$V$ curve of device A in zero magnetic field at the base temperature of our dilution refrigerator ($T$ = 8 mK) shows the junction critical current $I_{\rm c}$ = 0.39 $\mu$A, above which the junction shows a finite resistance (Fig. 3c). When the current $I_{\rm dc}$ is above 1.1 $\mu$A, the superconducting critical current of the PdTe$_2$ electrode is exceeded, which is marked by a horizontal jump in Fig. 3c. The jump occurs at the dc voltage $V_{\rm dc}$ = 0.11 mV which is much lower than the bias voltage $V_{\rm gap} \equiv 2\Delta/e \approx 3.5 k_B T_c /e \approx$ 0.27 mV necessary for reaching the linear $I$-$V$ region where the Andreev reflections no longer contribute to the transport across the junction \cite{Octavio1983}. For an SNS junction, such a linear $I$-$V$ region is important for the determination of the excess current \cite{Octavio1983, Flensberg1988}, which provides quantitative information on the junction transparency. 
Unfortunately, in device A, the superconductivity in the SC electrode is quenched by exceeding the critical current and the whole device was in the normal state to follow the Ohm's law, which can be judged from the fact that the linear $I$-$V$ relation at high bias extrapolates to the origin. 

Hence, estimation of the junction transparency through the excess current was not possible for device A. The $I$-$V$ curves showed no hysteretic behavior at any temperature, indicating that the junction was in the overdamped region. The overdamped nature of device A is understood by its small Stewart-McCumber parameter $\beta_{\rm c}  = 2\pi I_{\rm c} R_{\rm N}^{2} C /\Phi_0 \ll$ 1 \cite{Supplemental}.

Although the narrow SC electrode was problematic for the excess current measurement, it was necessary for observing the Fraunhofer pattern without being bothered by flux jumps. The mapping of d$V$/d$I$ in the $I_{\rm dc}$ vs $H$ plane is shown in Fig. 3d, where the dark blue region corresponds to the zero-resistance state and $I_{\rm c}$ can be inferred from its boundary in white or light blue colour. The quasi-periodic change in $I_{\rm c}$ with $H$ roughly follows a Fraunhofer-like pattern, but it deviates from the standard pattern described by $I_{\rm c}(H) = I_{\rm c0} |\sin(\pi \Phi_{\rm J}/\Phi_0)/(\pi \Phi_{\rm J}/\Phi_0)|$, where $I_{\rm c0}$ is the critical current at $H$ = 0,  $\Phi_{\rm J}$ is the magnetic flux threading the junction channel and $\Phi_0$ = $h/2e$ is the flux quantum. Our previous work on TI-based Josephson junctions explained similar deviations as a result of inhomogeneous current distributions in the junction area \cite{Ghatak2018}.

We emphasize that the observation of a Fraunhofer-like pattern gives evidence that the zero-resistance state of our junction was {\it not} caused by superconducting shorts between the SC electrodes. This is an important piece of information, because the PdTe$_2$ layer is self-formed beneath the deposited Pd electrode and it is not {\it a priori} clear how sharply the boundary of the PdTe$_2$ layer follow the boundary of the Pd electrode. Our data show that the boundary is at least sharp enough for fabricating a TI-based Josephson junction.

\subsection{SQUID}

\begin{figure*}[t]
\centering
\includegraphics[width=11cm]{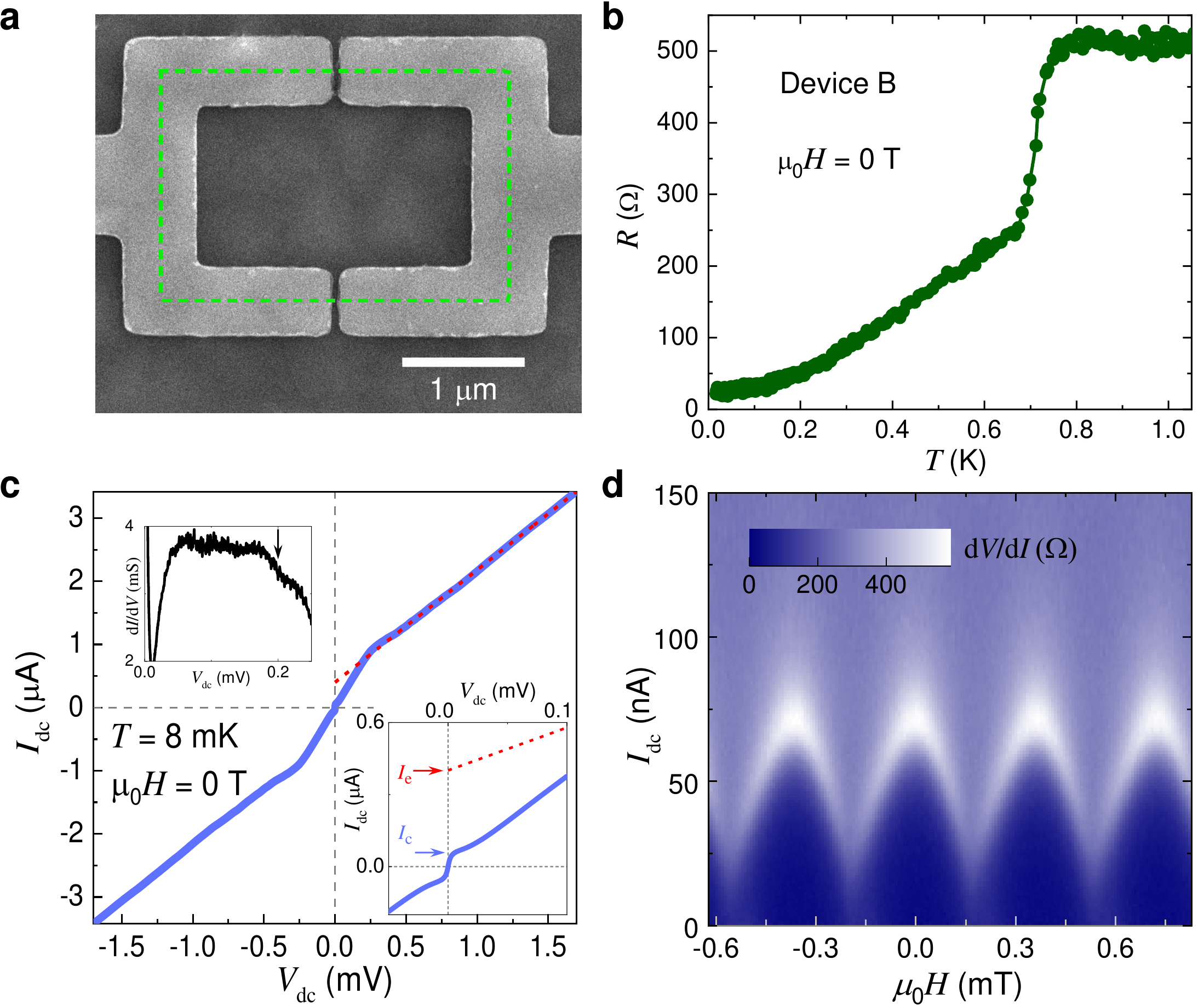}
\caption{SQUID fabricated by Pd deposition. (a) SEM image of the SQUID (device B) with Pd electrodes deposited on top of a 35-nm-thick (Bi$_{1-x}$Sb$_x$)$_2$Te$_3$ ($x$ = 0.85). The effective area of the SQUID illustrated by the green dashed square is $S_{\rm eff}$ = 6 $\mu$m$^2$. (b) Zero-bias resistance ($R$) of device B as a function of temperature in $H$ = 0. (c) $I$-$V$ characteristics of device B at $T$ = 8 mK in $H$ = 0.  The excess current $I_{\rm e}$ is estimated from the intercept of a linear fit to the high-bias region (red dashed line). Upper inset: $dI/dV$ vs $V_{\rm dc}$ for $0 \leq V_{\rm dc} \leq 0.25$ mV. Lower inset: zoom-in of the low-bias region of the $I$-$V$ characteristics. (d) 2D mapping of the deferential resistance (d$V$/d$I$) in the $I_{\rm dc}$ vs $H$ plane, measured at $T$ = 8 mK in device B. Clear periodic oscillations of the critical current (white bands) is characteristic of a SQUID. 
}
\label{fig:4}
\end{figure*}

To further test the usefulness of this self-epitaxy for superconducting devices, we have also fabricated a SQUID (device B). As shown in Fig. 4a, this device B has relatively wide ($\sim$1 $\mu$m) SC electrodes and two identical Josephson junctions with 50-nm length and 600-nm width. The effective area of the SQUID loop, $S_{\rm eff}$, was 6 $\mu$m$^2$. The SQUID design allowed us to measure the periodic change in $I_c(H)$ at very low fields, $\mu_0 H \lesssim$ 1 mT, where no flux jump occurs. 

The temperature dependence of the resistance of device B (Fig. 4b) shows that the PdTe$_2$ layer has $T_c \approx$ 0.70 K and there remains a finite apparent resistance ($\sim$ 25 $\Omega$) down to 8 mK, which is probably caused by noise, because the SQUID loop is much more susceptible to noise pick-up. In fact, device B showed a small but clear junction critical current (Fig. 4c inset), suggesting that the junctions were essentially in the zero-resistance state at the base temperature in zero field.
The mapping of d$V/{\rm d}I$ in the $I_{\rm dc}$ vs $H$ plane for device B (Fig. 4d) shows clean SQUID-type oscillations with the period $\mu_0 \Delta H$ = 0.36 mT, which is consistent with the expectation $\mu_0 \Delta H_{\rm cal} = \Phi_0/S_{\rm eff}$ = 0.34 mT. This $I_c(H)$ behavior confirms the successful formation of two Josephson junctions. 

The $I$-$V$ curve of device B at 8 mK (Fig. 4c) shows that the linear $I$-$V$ region at $V_{\rm dc} > V_{\rm gap} = 3.5 k_B T_c /e \approx$ 0.21 mV can be reached without driving the PdTe$_2$ layer out of superconductivity. This is expected from the wider SC electrodes than that in device A. In fact, in the plot of $dI/dV$ vs $V_{\rm dc}$ shown in Fig. 4c inset, a kink is seen at $\sim$0.2 mV, which is possibly associated with the primary Andreev reflection expected at $V_{\rm dc} = 2\Delta/e$, supporting the interpretation that the linear region above this $V_{\rm dc}$ reflects the physics without Andreev reflection. A linear fit was performed for this linear region and was extrapolated to $V_{\rm dc}$ = 0, yielding the excess current $I_{\rm e}$ = 0.40 $\mu$A, which is much larger than the critical current $I_{\rm c}$ = 0.07 $\mu$A of the SQUID. The normal-state resistance $R_{\rm N}$ = 562 $\Omega$ is also obtained from this linear fitting for the ``N'' part of the SNS junction; note that if the ``S'' part is driven into the normal state, the whole junction is in the normal state and the linear fitting to the $I$-$V$ relation extrapolates to the origin, as was the case with device A. In the data shown in Fig. 4c, the ``S'' part remains in the SC state, so that the $I$-$V$ relation allows us to extract a finite excess current.

Assuming that the two Josephson junctions are identical, the $I_{\rm e} R_{\rm N}$ product of each junction is calculated as $I_{\rm e}^{\rm JJ} R_{\rm N}^{\rm JJ} = (I_{\rm e}/2)(2R_{\rm N})= I_{\rm e} R_{\rm N} \approx$ 0.23 mV. Together with the superconducting gap estimated as $\Delta = 1.75k_B T_c \approx$ 0.105 meV, a barrier strength $ Z \approx$ 0.19 is obtained based on the Octavio-Tinkham-Blonder-Klapwijk theory \cite{Octavio1983, Flensberg1988}, which gives a junction transparency $T=1/(1+Z^{2})\approx$ 0.96. This is among the highest transparency reported for TI-based Josephson junctions \cite{Veldhorst2012, Oostinga2013, Galletti2014, Snelder2014, Ghatak2018, Jauregui2018, Schuffelgen2019}. 

\section{Conclusion}

The high junction transparency observed here is encouraging, but the lack of subgap features in the $dI/dV$-vs-$V$ relation to signify the existence of multiple Andreev reflections, which is usually observed in high-transparency SNS junctions \cite{Ghatak2018,Jauregui2018}, tells us that one should be cautious about concluding the nature of the junctions. Since the main purpose of the present paper is to report the discovery of unexpected self-epitaxy to form a superconductor on the surface of BST, we leave it for future studies to understand the details of the self-formed superconducting materials as well as the characterization of the interface to the TI surface states. Epitaxial growth of a superconductor on the TI surface is rare, and to our knowledge, the MBE-growth of PdTe$_2$ on Bi$_2$Te$_3$ by Xue {\it et al.} \cite{Xue2019} is so far the only report, although in that work no device was made and the superconducting property of PdTe$_2$ was poorly characterized. The inverted scheme, i.e. epitaxial growth of a TI on the surface of a superconductor, have been tried in the past \cite{Wang2012, Wang2013, Xu2014}, but the fabrication of a Josephson junction is impossible in such a structure and the characterizations have been performed only via scanning tunneling microscopy or angle-resolved photoemission spectroscopy. Therefore, the new and convenient scheme for preparing an epitaxial SC/TI interface compatible with nano-device fabrication is a useful step forward in the realization of the Fu-Kane proposal in TIs to engineer a TSC state. 

\acknowledgements{
We thank Max Knuth for helping the preparation of STEM samples. This project has received funding from the European Research Council (ERC) under the European Union's Horizon 2020 research and innovation programme (grant agreement No 741121) and was also funded by the Deutsche Forschungsgemeinschaft (DFG, German Research Foundation) under CRC 1238 - 277146847 (Subprojects A04 and B01) as well as under Germany's Excellence Strategy - Cluster of Excellence Matter and Light for Quantum Computing (ML4Q) EXC 2004/1 - 390534769. G.L. acknowledges the support by the Research Foundation - Flanders (FWO, Belgium), file nr. 27531 and 52751.}

\clearpage
\onecolumngrid

\renewcommand{\thefigure}{S\arabic{figure}} 
\renewcommand{\thetable}{S\arabic{table}} 

\setcounter{figure}{0}
\setcounter{section}{0}

\begin{flushleft} 
{\Large {\bf Supplemental Material}}
\end{flushleft} 
\vspace{2mm}

\section{AFM measurements before and after P\MakeLowercase{d} deposition}

The morphology of sample 1 was characterized before and after the Pd deposition using an atomic force microscope (AFM), as shown in Fig. S1. The high quality of the as-grown (Bi$_{1-x}$Sb$_x$)$_2$Te$_3$ (BST) film is confirmed by the large, atomically flat terraces seen on the surface. After the deposition of a 8-nm Pd layer by magnetron sputtering, the triangular structures of the BST film beneath are still visible (Fig. S1\textbf{b}), indicating that the sputtered Pd layer covers the BST surface uniformly. The sputtered Pd layer forms a polycrystalline film. Nano-grains of Pd with a mean diameter of $\sim$7 nm are seen in the small-area scanning, as shown in Fig. S1\textbf{c}.

\begin{figure}[h]
\centering
\includegraphics[width=0.85\textwidth]{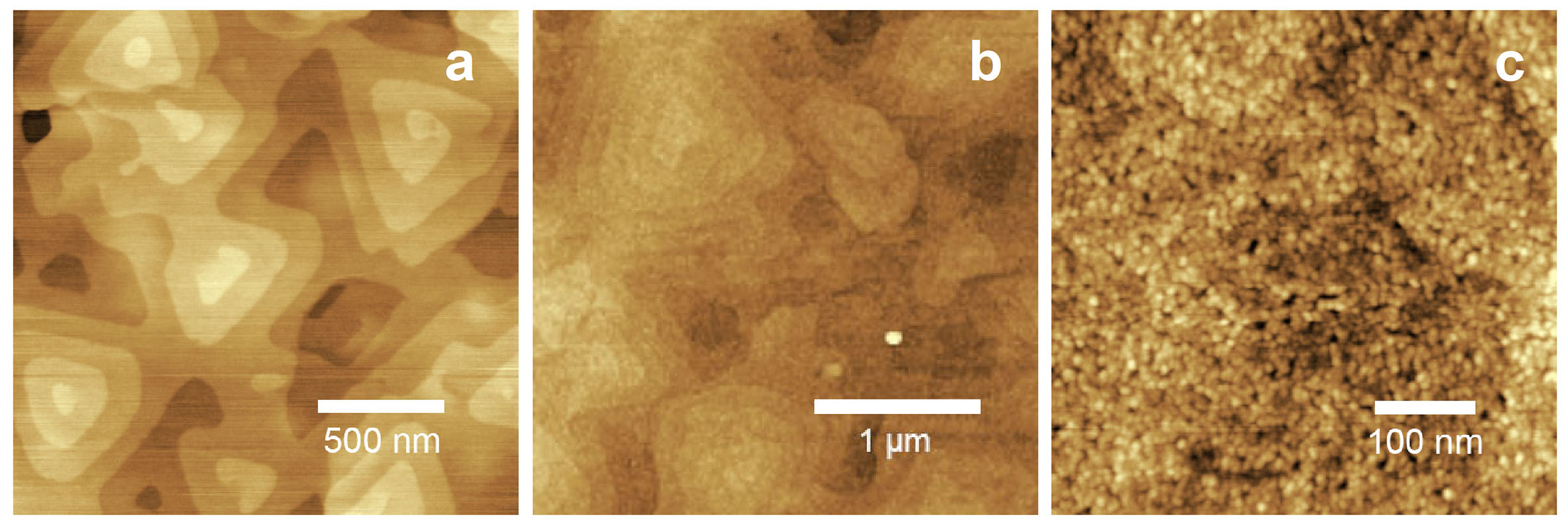}
\caption{AFM images of sample 1. (a) AFM image of an as-grown (Bi$_{1-x}$Sb$_x$)$_2$Te$_3$ film. (b,c) AFM images of sample 1 after Pd deposition in different magnifications.}
\label{fig:s2}
\end{figure}

\section{Additional STEM data}

A high-resolution scanning transmission electron microscopy (STEM) image of the epitaxial PdTe$_2$ layer is shown in Fig. S2\textbf{a}. The Te-Pd-Te triple-layers of PdTe$_2$ are clearly resolved, as indicated by the color balls superimposed on the STEM image. The outer tellurium atoms are heavier than the inner palladium atoms and therefore are larger and brighter in the STEM image. Some atoms (most likely Bi or Sb) are occasionally seen intercalated in the van der Waals gap of PdTe$_2$; an example of an intercalated atom is pointed by the arrow in Fig. S2\textbf{a}.

The lattice constants of bulk PdTe$_2$ are $a$ = 4.04 \si{\angstrom} and $c$ = 5.13 \si{\angstrom} \cite{Liu2018}. To obtain the lattice constants of the self-formed PdTe$_2$ layer, two-dimensional fast Fourier transform (2D FFT) was performed on a high-resolution STEM image of the PdTe$_2$ phase (Fig. S2{\bf b}). The lattice constants obtained from the FFT analysis are $a$ = 4.07 \si{\angstrom} and $c$ = 5.44 \si{\angstrom}, corresponding to a lattice expansion of 6.0\% in the $c$ direction compared to the bulk value. Such an expansion is larger than the value of 2.1\% reported for a 10-triple-layer-thick PdTe$_2$ film grown on SrTiO$_3$ (001) surface \cite{Liu2018}. The relatively large lattice expansion in the $c$ direction is most likely due to the intercalation of other atoms in the van der Waals gap.

\begin{figure}[h]
\centering
\includegraphics[width=0.9\textwidth]{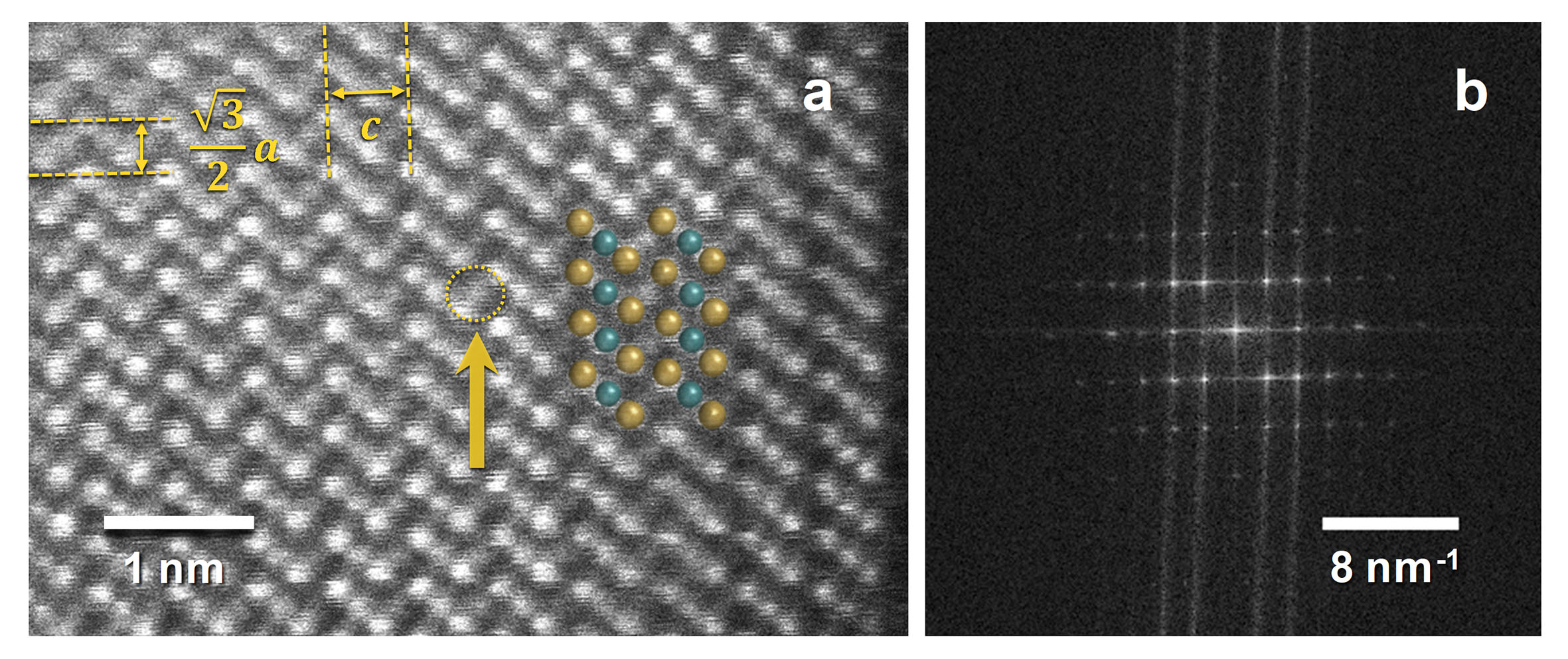}
\caption{Additional STEM data of the self-formed PdTe$_2$ phase. (a) High-resolution STEM image of the PdTe$_2$ layer, taken at $V_{\textrm{acc}}=200$ kV. The Te and Pd atoms are indicated by the superimposed balls in dark yellow and dark turquoise colour, respectively. The thick arrow points to an atom intercalated in the van der Waals gap of PdTe$_2$. (b) The 2D FFT image of a STEM picture of the PdTe$_2$ layer.}
\label{fig:s2}
\end{figure}

\section{Estimation of the Stewart-M\MakeLowercase{c}Cumber parameter}

The $I$-$V$ characteristics of device A show no hysteresis, as seen in Fig. 3\textbf{c} of the main text. According to the RCSJ model of Josephson junctions, the absence of hysteresis in $I$-$V$ curves indicates an Stewart-McCumber parameter $\beta_{c}~=~2\pi$$I_{\textrm{c}}$$R_{\text{N}}^2$$C$/$\Phi_0\ll1$, where $\Phi_0$ is the flux quantum, $C$ is the shunt capacitance, $I_{\textrm{c}}=391$ nA and $R_{\text{N}}=110$ $\Omega$ are the critical current and the normal-state resistance of the Josephson junction, respectively. Here $R_{\text{N}}$ is approximated by the junction resistance right after the major superconducting transition in the cooling curve, as plotted in Fig. 3\textbf{a} of the main text. 
The shunt capacitance $C$ of device A mainly comes from the capacitive coupling between the Pd electrodes, and its order of magnitude can be roughly estimated using the formula for parallel-plate capacitor, which gives $C=\epsilon_0wd/l\sim10^{-17}$ F, where $d=30$ nm is the thickness of the Pd electrodes, $w=3$ $\mu$m and $l=100$ nm are the width and length of the Josephson junction, respectively. Note that, due to the small $d/l$ ratio, the actual value of $C$ would be larger than the value estimated using the parallel-plate capacitor formula, but they are of the same order.

With these values of $I_\textrm{c}$, $R_\textrm{N}$ and $C$, the Stewart-McCumber parameter of device A is indeed estimated to be $\beta_{c} \sim 10^{-4}\ll1$, which is consistent with the observed non-hysteretic behaviour of the $I$-$V$ curves.

\normalsize

\end{document}